\documentclass[12pt]{article}
\usepackage{epsfig,graphicx,color}
\usepackage{amsmath}
\usepackage{latexsym}
\usepackage{amstext}
\usepackage{epsfig, graphicx}
\newcommand{\ds}{\displaystyle}
\newcommand{\beq}{\begin{eqnarray}}
\newcommand{\eeq}{\end{eqnarray}}
\newcommand{\beqq}{\begin{eqnarray*}}
\newcommand{\eeqq}{\end{eqnarray*}}

\newcommand{\eps}{\varepsilon}

\font\bb=msbm10 at 12pt

\def\eE{\hbox{\bb E}}
\begin{document}
\pagestyle{plain}
\begin{center}
{\large \textbf{{Do cells sense time by number of divisions?}}}\\[5mm]
  {Z. Schuss\footnote{Tel-Aviv University, Tel-Aviv 69978, Israel.}, K. Tor and D.
Holcman\footnote{  Ecole Normale Sup\'erieure, 46 rue d'Ulm 75005 Paris, France. This research was
supported by a FRM team grant. }}
\end{center}
\date{}
\begin{abstract}
Do biological cells sense time by the number of their divisions, a process that
ends at senescence? We consider the question "can the cell's perception of time be expressed through the length of the shortest telomere?" The answer is that the absolute time before senescence cannot be expressed by the telomere's length and that a cell can survive many more divisions than intuitively
expected. This apparent paradox is due to shortening and elongation of the telomere, which suggests a random walk model of the telomere's length. The model indicates two phases, first, a determinist drift of the length toward a quasi-equilibrium state, and second, persistence of the length near an attracting state for the majority of divisions prior to senescence. The measure of stability of the latter phase is the expected number of divisions at the attractor (''lifetime") prior to crossing a threshold to senescence. The telomerase regulates stability  by creating an effective potential barrier that separates statistically the shortest lifetime from the next shortest. The random walk has to overcome the barrier in order to extend the range of the first regime. The model explains how random telomere dynamics underlies
the extension of cell survival time.
\end{abstract}

\section{Introduction}
Finding a measure of our sensation of time is an intriguing question in physical and life sciences. In cell
biology, the lifetime of a cell is reflected in the number of cell divisions prior to senescence, which is a measure of a cell's lineage death and is a component of cellular aging \cite{Aubert}. The number of cell divisions is expressed by the length of telomeres, which protect the ends of the chromosomes. Telomeres can lose between a few to hundreds of base pairs during cell division, or increase their length through the action of the enzyme telomerase. We are concerned here with the physical mechanism that regulates the number cell of divisions prior to senescence.

Several decades of research have revealed that telomeres \footnote{Telomeres are sections
of DNA, found at the ends of each chromosome. They consist of the same sequence of bases repeated
over and over. In humans the telomere sequence is TTAGGG. This sequence is usually repeated about
3,000 times and can reach up to 15,000 base pairs in length (see
https://www.yourgenome.org/facts/what-is-a-telomere. Senescence means biological aging, telomerase is an enzyme that extends the telomeres of chromosomes
\cite{Cech,Texiera,Hemann,Tan,Vaugh,Charchar}).}  are made of repetitive nucleotide sequences at each end of a chromatid, which protects the end of the chromosome from deterioration or from fusion with other chromosomes. Following each cell division, the telomere ends become shorter on the average  \cite{Wikipedia-telomere}. Telomeres are necessary for the maintenance of chromosomal integrity and overall genomic stability \cite{delange,Marcand} and in the absence of any mechanism of elongation, telomere length can only decrease over time \cite{Watson}. As a result, a cell can divide only a finite number of times
before proliferation is arrested . However, there is a mechanism that inhibits shortening and  is due
to the action of an enzyme (telomerase) that can elongate telomeres, thus inhibiting their monotonic decrease to a critical length, which defines the threshold of senescence. Interestingly, short telomeres are preferentially elongated by telomerase \cite{Texiera} and the shortest telomere is apparently a limiting factor of cellular proliferation \cite{Hemann}.

It is not yet possible to monitor telomere dynamics throughout cell division, so that theoretical
models have been used to predict telomere shortening under various conditions: models have
revealed the molecular dynamics and the variability of triggering senescence in mammalian cells
within a telomerase-deficient cell population  \cite{Tan,Proctor1,Proctor2,Buijs,Rodriguez}.
More recently, we developed a stochastic model, based on an asymmetric random walk, which
revealed that in a steady-state population of telomeres, there is a statistical gap
between the shortest and the remaining telomere lengths, suggesting that the shortest one can
play a key role in determining the number of cell divisions and the trigger of senescence
\cite{Redner,Zhou,Daoduc,Viewpoint}.

Because the length of the telomere does not strictly reflect  the number of cell divisions, we
focus here on the time-dependent telomere dynamics prior to senescence in order to determine the
statistics of the number of cell divisions prior to senescence. The expected number of divisions
of the majority of cells, beginning with yeast and up to human immune or reproductive cells, is about 70--100, while the shortening process allows the cell to survive only about 25--50 divisions. Our model
suggests an answer to the question "where are the missing divisions?" Normally, cells  can divide
only about 50 to 70 times, with telomeres getting progressively shorter, until the cells become
senescent, dies, or sustains genetic damage that could cause cancer. The known expected number of
divisions of a human reproductive cell prior to senescence is about 70, while the shortening
process allows the cell to survive only about 25 divisions.  Similarly, elimination of telomerase
in budding yeast leads to a decline in viability, leading to a colony that can survive 50
generations compared to the normal colony that can survive 75 to 100 generations \cite{Ballew}.

To answer the above question and to explain the stopover on the way to senescence following telomerase elimination, both qualitatively and quantitatively, we analyze and simulate the extreme statistics of an asymmetric random-walk model of telomere length. We find that the random walk has a quasi steady-state, in which the telomere is sufficiently short for the telomerase-induced elongation to offset the division-induced shortening of the telomere. Interestingly, telomeres spend most of their lifetime in this state prior to senescence. We distinguish between two stages in the arrival to senescence, the shorter period prior to reaching the above-mentioned state (of quasi-equilibrium) and the longer time spent there prior to senescence. We conclude that the length of all telomeres, in particular the shorter ones, do not reflect the progressive decline in the number of the remaining cell divisions and, surprisingly, the extension of the number of cell divisions is a manifestation of the stochastic process of telomere
elongation and shortening.

\section{Results}
\subsection{A model of telomere dynamics}
We adopt the random walk model of telomere dynamics \cite{Zhou,Daoduc}, in which
the length $x$ of the telomere can decrease or increase in each division. The model assumes that
the length decreases by a fixed length $a$ with probability $l(x)$ or, if recognized by a
polymerase, it
increases by a fixed length $b\gg a$ with probability $r(x)=1-l(x)$. The jump probability
$r(x)$ is assumed a decreasing function of $x$ with $r(0)=1$ \cite{Daoduc}. Thus the length of
the telomere at the $n$-th division is an asymmetric random walk $x(n)$. In our simplified model,
we assume that the maximal length of a telomere is $L\gg b$. We assume that senescence ensues when the length decreases below a critical value $T$ , that is, the division process stops.

The problem at hand is to determine the evolution of the telomere length, to study the dynamics
of the shortest length, and to investigate the role of the probability $r(x)$, which can be
modulated by telomere diseases \cite{Armanios}. In particular, we study the statistics of the
trajectories $x(n)$, their expected time to reach their quasi-stationary state, and the expected  number of divisions before reaching the threshold $T$ for the first time.
\subsection{The asymmetric random walk model}
\begin{figure}[http!]
\includegraphics[width=0.32\textwidth]{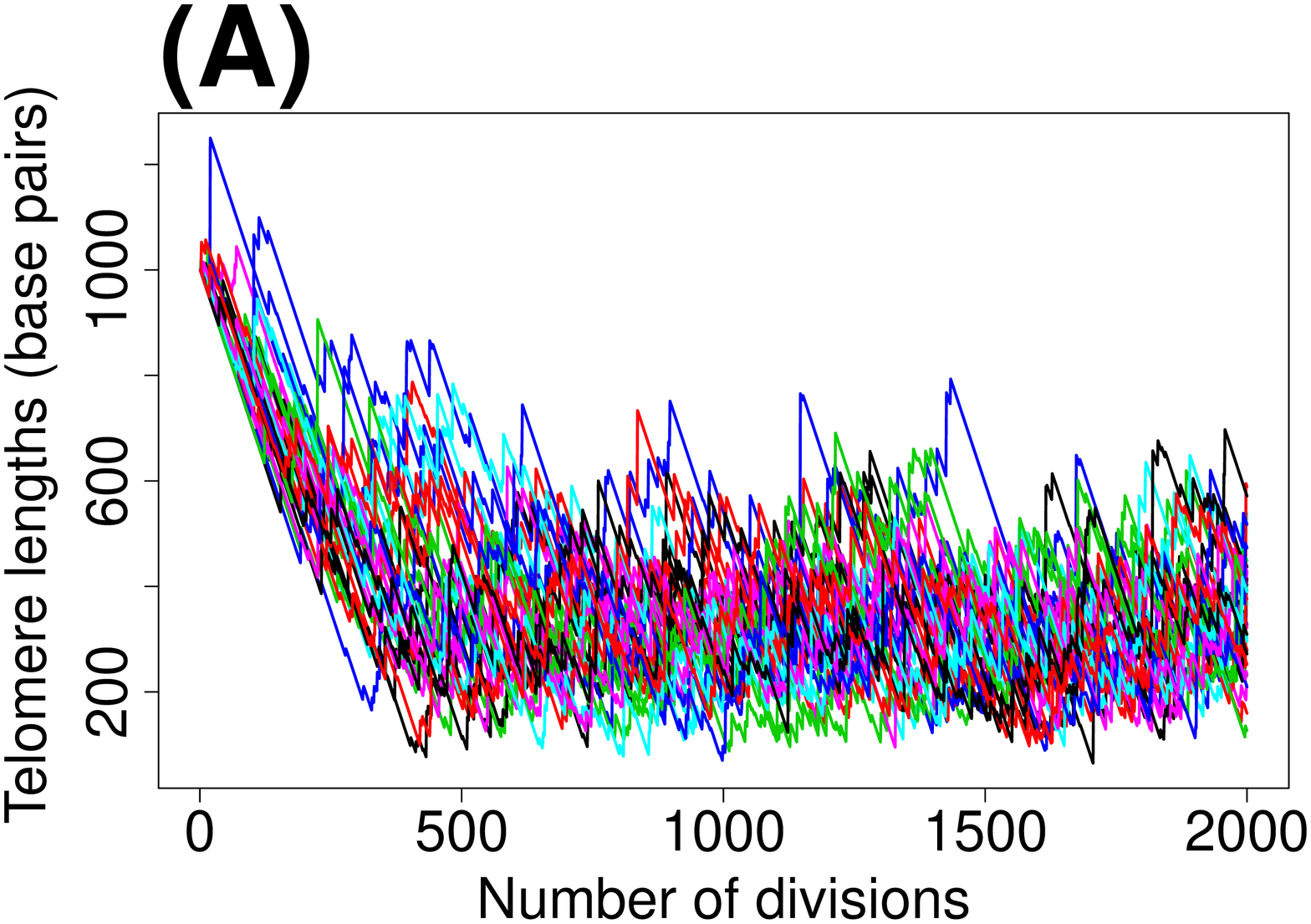} 
\includegraphics[width=0.32\textwidth]{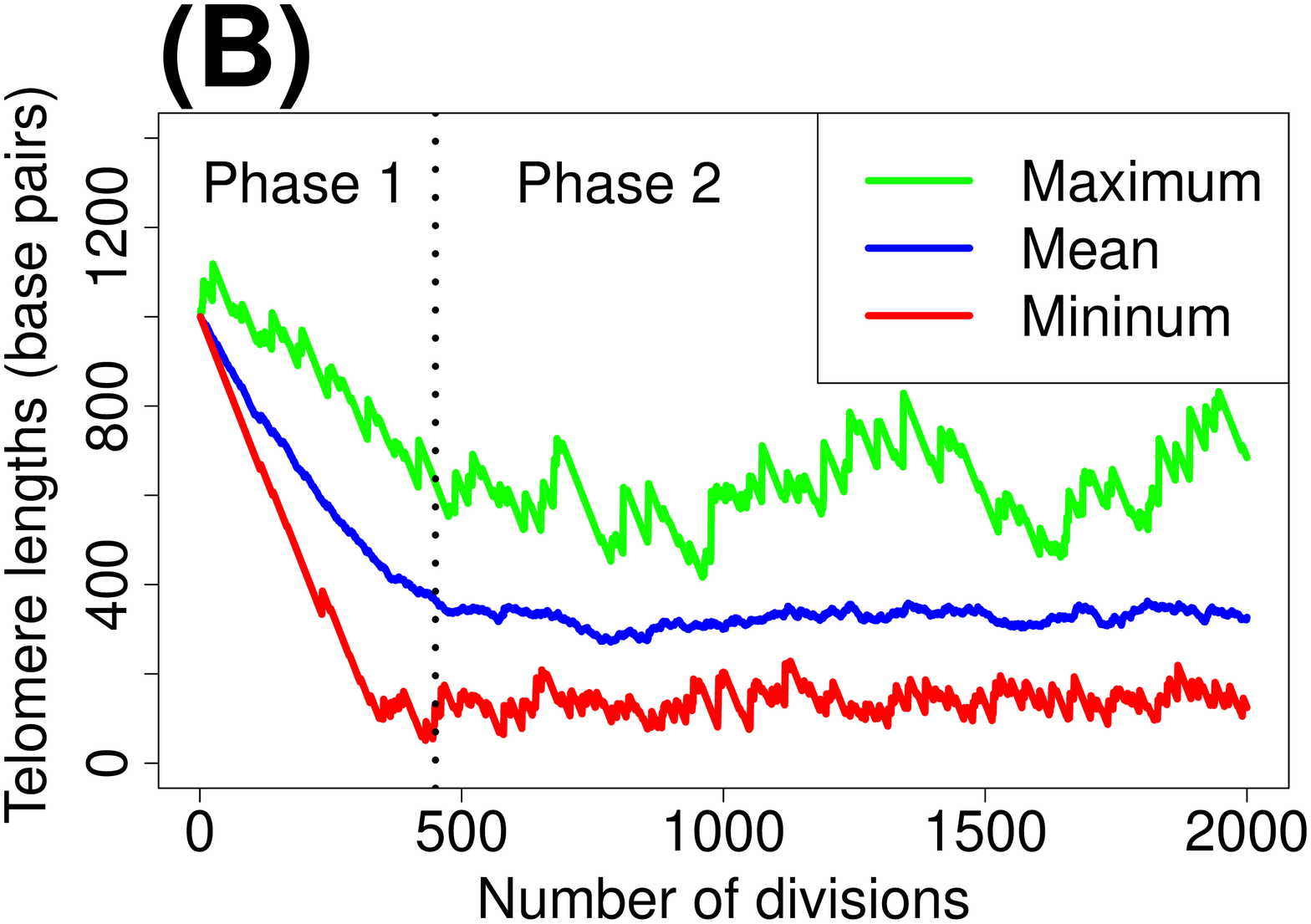}
\includegraphics[width=0.32\textwidth]{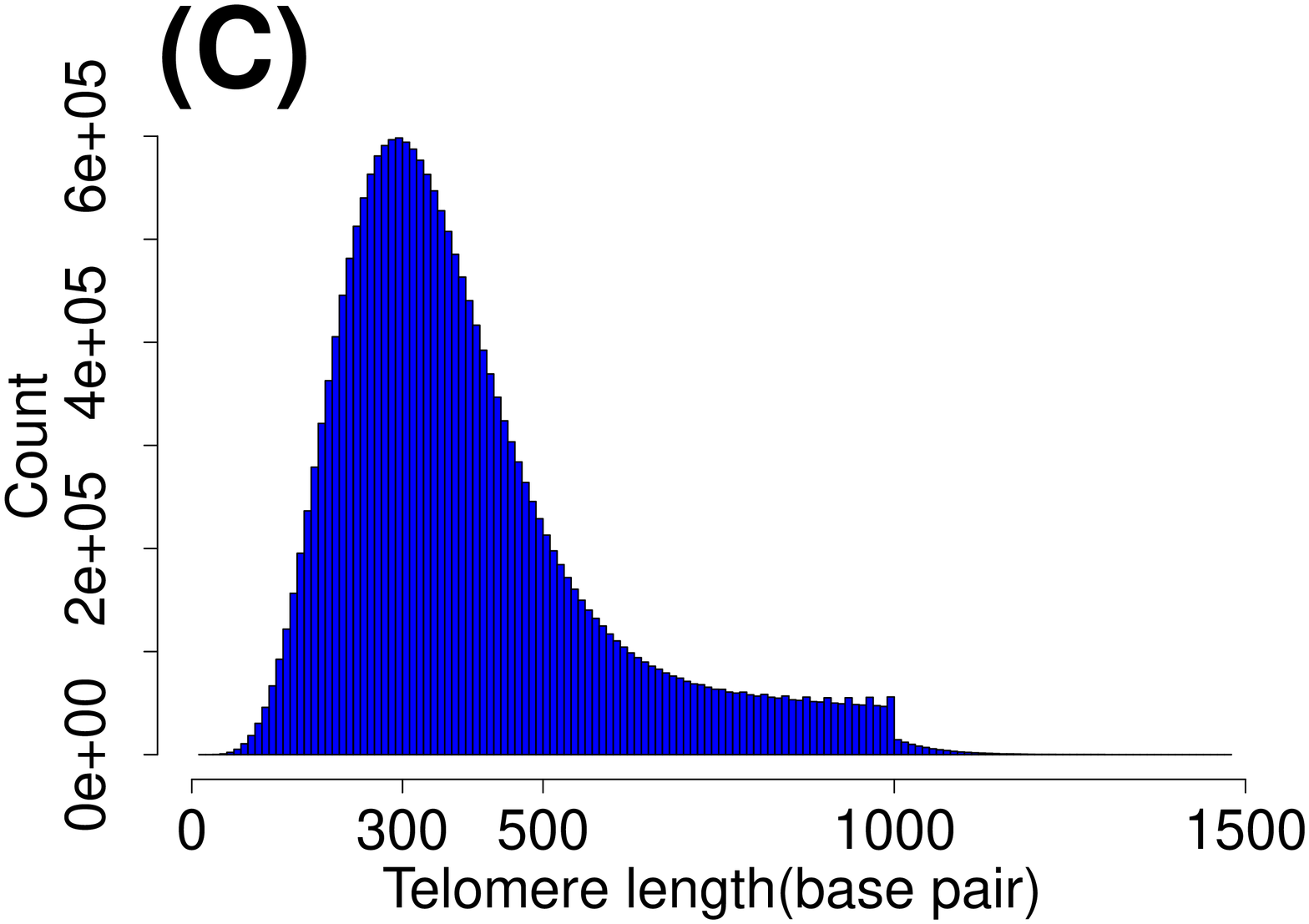}
\caption{\small{\bf Telomere dynamics in {yeast.}} \textit{\bf (A)} Stochastic dynamics of the 32 telomeres (see \eqref{dynamics} with elongation exponentially distributed with mean 40 bps.) {\bf (B)} Dynamics of the minimum, maximum, and the mean. Two phases are seen: phase 1 is characterized by constant decay and convergence toward phase 2, which is a quasi steady-state due to an attractor, which prevents the collapse of the telomere to a critical value. There is a large increase in the number of cell divisions in phase 2. \textit{\bf (C)} Distribution of telomere length at steady-state. The parameters are $\beta=0.045$ and $a=3\pm 1/2$.}
\label{f:figure1}
\end{figure}

The model of the telomere dynamics is
\begin{align}x_{n+1}=\left\{\begin{array}{ll}x_n-a&\mbox{w.p.}\
l(x_n)\\ \\
x_n+b&\mbox{w.p.}\ r(x_n),\end{array}\right.\label{dynamics}
\end{align}
where the right-probability $r(x)$ can be approximated by
\begin{align}
r(x)=\frac{1}{1+\beta x},\label{rx}
\end{align}
for some $\beta>0$. The model \eqref{dynamics} is simulated for all telomeres (with a maximum of
32), and the statistics of the average, the longest, and the shortest trajectories are calculated.
The representative parameters are as follows. The shortening law is $a=3$ and the elongation law
is $b=30$. The results are quite similar to an exponential law with mean $b=30$ (see
\cite{Daoduc}). The results shown in Fig.\ref{f:figure1} reveal two phases: in the first one, the
telomere's length decreases almost deterministically to a quasi-equilibrium length as
described below. In the second phase, the length persists in the quasi-equilibrium state for the
majority of divisions. In the absence of any stopping process, the telomere's length stays near
its minimum, so apparently the cell can live forever.
\subsection{Two phases and the quasi-equilibrium state}
To characterize each phase, we use the scaling $x_n=y_nL$ and setting $\eps=b/L$, the dynamics \eqref{dynamics} becomes
\begin{align}
y_{n+1}=\left\{\begin{array}{ll}y_n- \ds\frac{\eps a}{b}&\mbox{w.p.}\
\tilde l(y_n)\\
y_n+\eps &\mbox{w.p.}\ \tilde
r(y_n),\end{array}\right.\label{tdynamics}
\end{align}
where $\tilde l(y)=l(x)$. In the limit $\eps\ll1$, the process $y_n$ moves in small steps. The dynamics (\ref{tdynamics}) falls under the general scheme \cite{KMST1,KMST2,KMST3,Schuss}
\begin{equation}
y_{n+1}=y_n+\eps \xi _n,  \label{2RW}
\end{equation}
where
\begin{align}
\Pr\left\{ \xi _n=\xi \,|\,y_n=y,\,y_{n-1}=y_1,\,\ldots\right\}
=w(\xi\,|\,y,\eps),\label{Pxi}
\end{align}
$\eps$ is a small parameter, and $y_0$ is a random variable \index{random variable} with a given pdf $p_0(y)$. In the case at hand the function $w(\xi\,|\,y)$ defined in (\ref{Pxi}) is given by
\begin{align}
w(\xi\,|\,y)=(1-\tilde
r(y))\delta\left(\xi+\frac{a}{b}\right)+\tilde r(y)\delta(\xi-1),\label{wxiy}
\end{align}
so the conditional jump moments are
\[m_n(y)=\left(-\frac{a}{b}\right)^n(1-\tilde
r(y))+\tilde r(y).
\]
The probability density function (pdf) of $y_n$ satisfies the forward master equation
\begin{align}
p_{\eps}(y,n+1\,|\,x,m)=p_{\eps}\left(y+\eps\frac{a}{b},n\,|\,x,m\right)\tilde
l\left(y+\eps\frac{a}{b}\right)+p_{\eps}(y-\eps,n\,|\,x,m)\tilde
r(y-\eps)\label{MA}
\end{align}
and the backward equation
 \begin{align}
&\, p_{\eps}(y,n\,|\,x,m)-p_{\eps}(y,n\,|\,x,m+1)\label{2BKE}\\
=&\,p_{\eps}\left(y,n\,|\,x-\eps\frac{a}{b},m+1\right)(1-\tilde
r(x))+p_{\eps}(y,n\,|\,x+\eps,m+1)\tilde r(x)-p_{\eps}(y,n\,|\,x,m).\nonumber
 \end{align}
The first conditional jump moment,
\begin{align}
m_1(y)=- \frac{a\tilde l(y)}{b}+\tilde r(y),\label{m1}
\end{align}
 changes sign at
\beq
z_0=\frac{b}{L\beta a}.\label{z0}
\eeq
If the process is terminated at the threshold $T$ mentioned above, then $\eps T/b<y<1$. In this case, the pdf  $p_{\eps}(y,n)$ converges to a quasi-stationary density $p_{\eps}(y)$ for large
$n$, prior to the termination of the trajectory $y_n$ at $y=T/L$. The quasi-stationary master
equation (\ref{MA}) becomes
\beq
p_{\eps}(y)=p_{\eps}\left(y+\eps\frac{a}{b}\right)\tilde
l\left(y+\eps\frac{a}{b}\right)+p_{\eps}\left(y-\eps \right)\tilde
r\left(y-\eps\right),\label{stMA}
\eeq
which for small $\eps$ is peaked near $z_0$, as shown in fig. \ref{f:figure1}C and \ref{f:figure2}A.
Indeed, to find the asymptotic structure of the quasi-stationary solution $p_{\eps}(y)$ for
$\eps\ll1$, we construct its approximation in the WKB form \cite{Schuss}
$p_{\eps}(y)=K_\eps(y)\exp\left\{-\psi(y)/\eps\right\}$, where $\psi(y)$ is the solution
of the eikonal equation
\beq
\tilde l(y)\exp\left\{-\psi'(y)\frac{a}{b}\right\}+\tilde
r(y)\exp\left\{\psi'(y)\right\}=1.\label{eikonaleq}
\eeq
The derivatives of $\psi(y)$ at $z_0$  shows a single peak near $z_0$ . Indeed $\psi'(z_0)=0$  and
\beq
\psi''(z_0)&=&\frac{2aL\beta}{a+b},\\
\psi'''(z_0)&=&-\frac43\frac{(a+2b)\beta^2L^2a^2}{b(a+b)^2}.
\eeq
\subsection{The expected lifetime of phase 1}
The essence of the mentioned paradox is due to the behavior of the first conditional moment $m_1(y)$ of the jump size (see \eqref{m1}). Because $m_1(z_0)=0$ and $m'_1(z_0)<0$ the random walk
drifts from any initial state $y$, for example from $y=1$, toward $y=z_0$, where it is
trapped in quasi-equilibrium fluctuations about $z_0$ for an expected number of jumps
$\bar n$, which may be larger than the expected number of jumps $n_{y\to z_0}$ that is
required to reach $z_0$ from $y$ for the first time. Specifically, the expected number
of jumps {(expected lifetime)} $n_{z_0\to T}$ to go from $z_0$ over the threshold (or 0) may be larger than $n_{1\to z_0}$.

To study this trapping phenomenon, we note first that the expected lifetime $n_0(y)$ to cross the boundary $y=z_0$ from an internal point $z_0<y<1$ is the solution of  \cite{Schuss}{
\begin{align}
{\cal L}_nn_0=\int\limits_{(z_0-y)/\eps}^{(1-y)/\eps} n_0(y+\eps
\xi)w(\xi\,|\,y)\,d\xi-n_0(y)=-1\hspace{0.5em}\mbox{for}\ z_0\leq y\leq1,\ \
n_0(y)=0\hspace{0.5em}\mbox{for}\ y<z_0.\label{BMEz0}
\end{align}
Setting $\tau(y)=\eps n_0(y)$, we write the Kramers-Moyal expansion of (\ref{BMEz0}) as
\begin{align}
\sum_{k=1}^\infty\frac{\eps^{k-1}m_k(y)}{k!}\frac{d^k\tau(y)}{dy^k}=-1\hspace{0.5em}\mbox{for}\
z_0\leq y\leq 1,\ \ \tau(z_0)=0,\quad \tau(1)=0, \label{KMz0}
\end{align}
and the diffusion approximation to (\ref{KMz0}) is
\beq
{\cal L}_\eps\tau_0(y)=\frac{\eps}{2}m_2(y)\tau_0''(y)+m_1(y)\tau_0'(y)=-1\hspace{0.5em}\mbox{for}\
z_0\leq y\leq 1,\quad \tau_0(z_0)=0,\ \tau'_0(1)=0.\label{diffKME}
\eeq
To show that the solution of (\ref{diffKME}) is a valid approximation to that of
\eqref{KMz0} it suffices to show that $\tau(y)$ has a convergent series representation
\beq
\tau(y)=\sum\limits_{k=0}^\infty\eps^k\tau_k(y),
\eeq
where $\tau_k(y)$ are bounded functions. The stochastic dynamics corresponding to \eqref{diffKME} is
\beq
\dot{y}=m_1(y)+\sqrt{\eps m_2(y)} \dot{w}(t),\label{DA}
\eeq
where $\dot w(t)$ is $\delta$-correlated Gaussian white noise and $t$ is the smoothly interpolated $\eps n$. In particular, the effective potential well is
\beq
V(y)=-\frac{2}{\eps}\int\limits_{y}^1\frac{m_1(z)}{m_2(z)}\,dz=(1-y)\frac{2b}{\eps a}
+\frac{2(a+b)b}{La^3\beta}\log\frac{a^2\beta y+\eps b}{a^2\beta z_0+\eps b},
\eeq
for which the drift $-V'(y)$ vanishes at $z_0$. Fig.2A  shows the effective potential $V(y)$ for
various values of the parameter $\beta$. Changing $\beta$ affects both the height and the location of the minimum of $V(y)$}.
{
\begin{figure}[http!]
\includegraphics[width=0.5\textwidth]{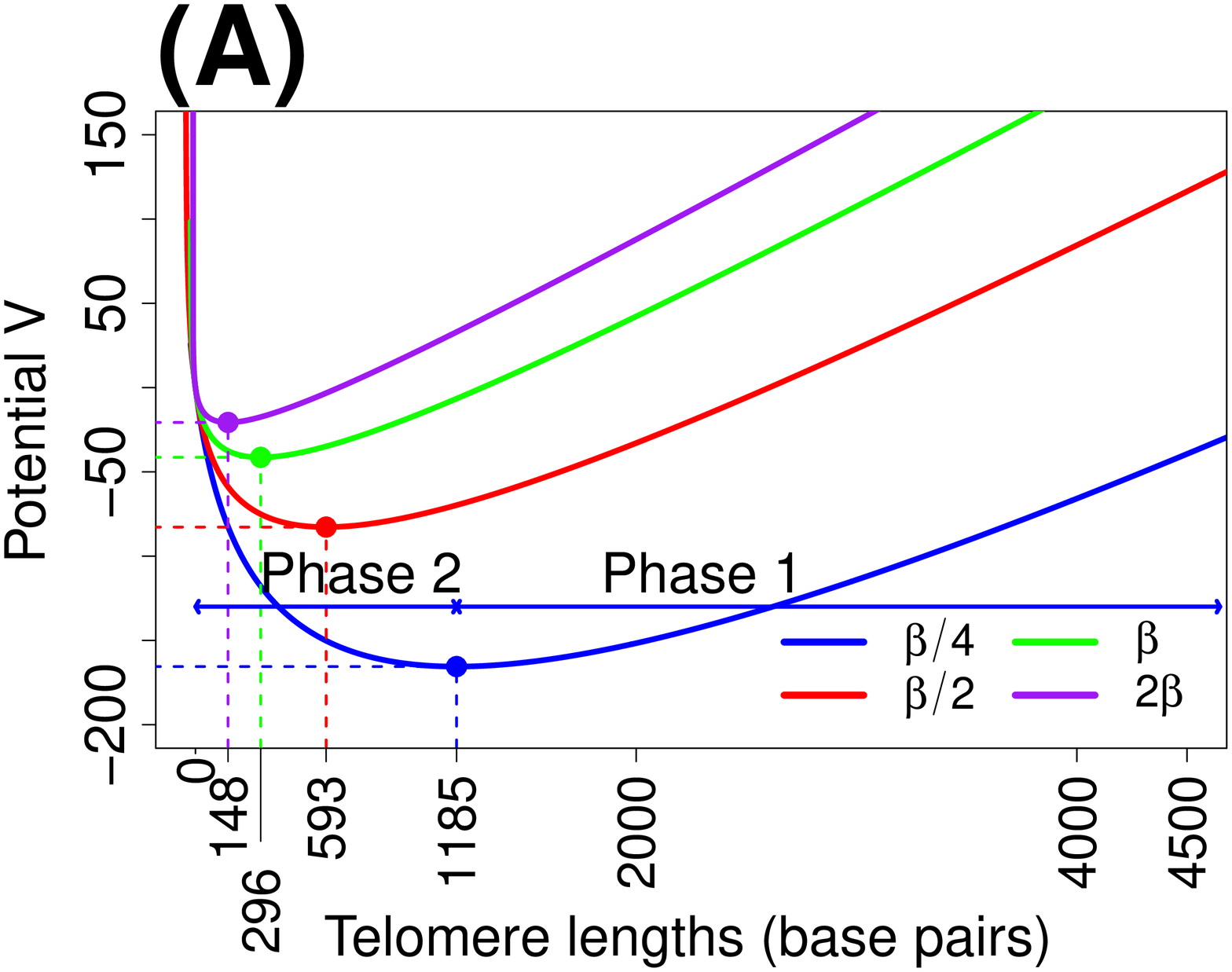} 
\includegraphics[width=0.5\textwidth]{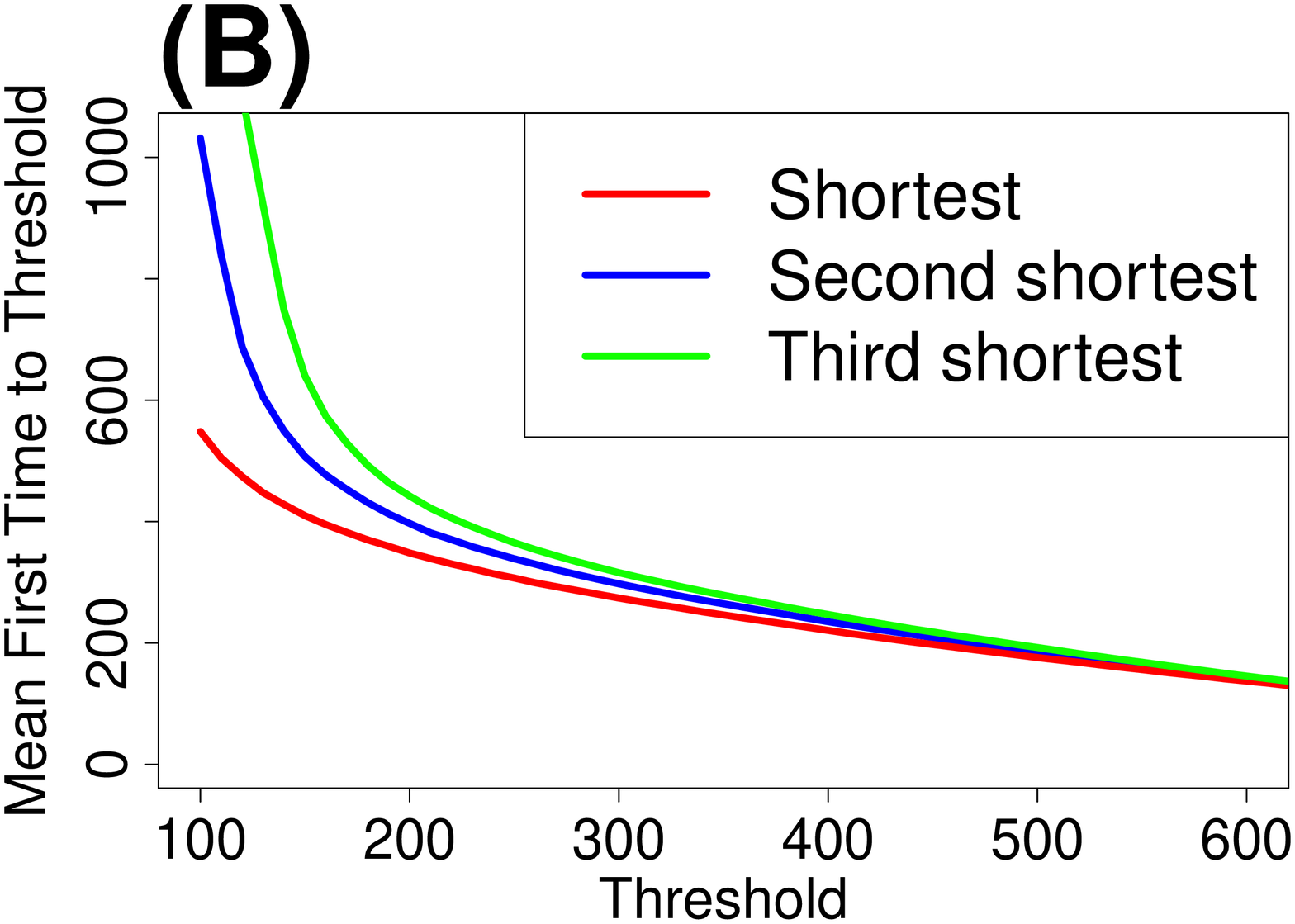}
\includegraphics[width=0.5\textwidth]{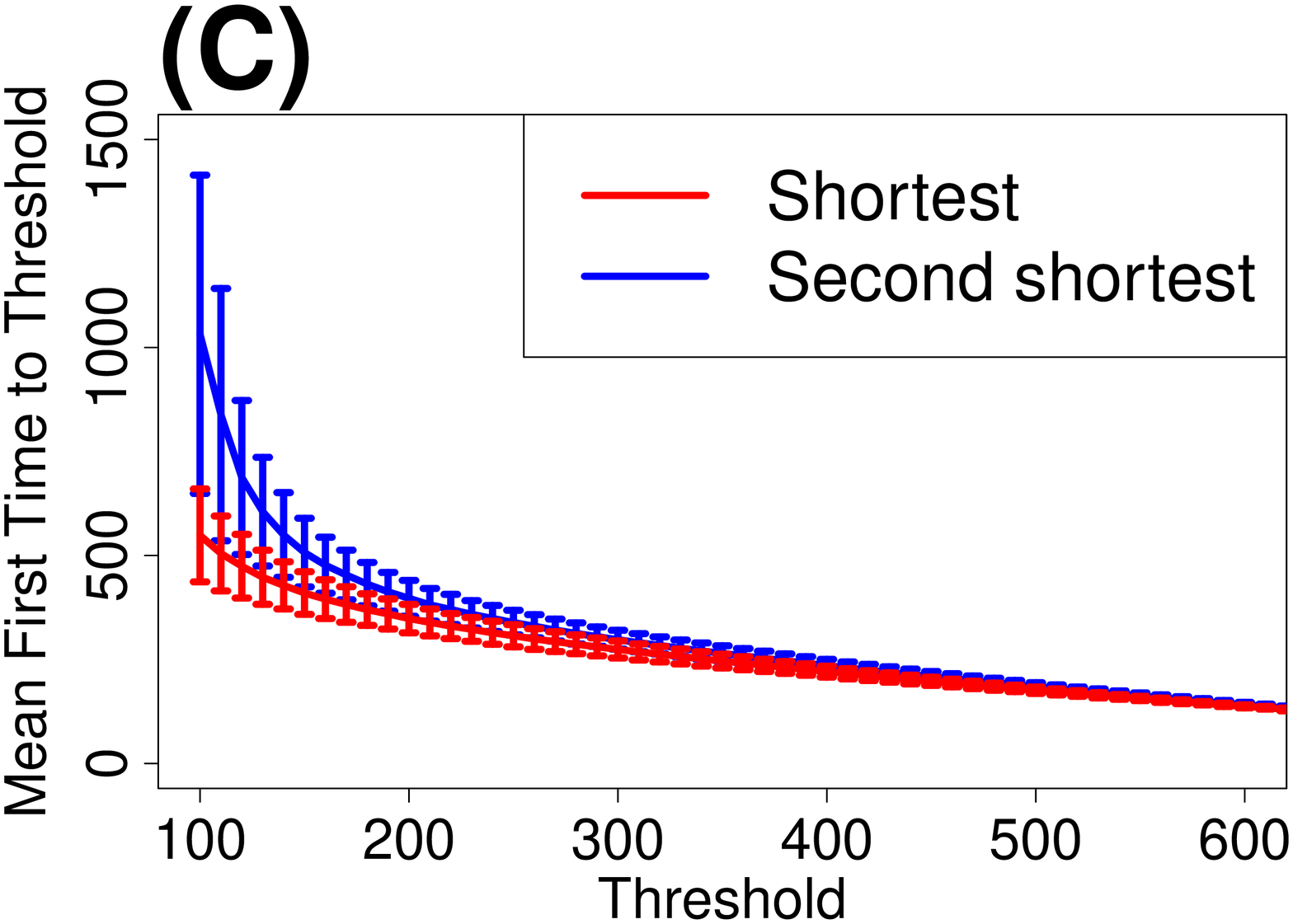} 
\caption{\small{\bf Statistical properties of telomere dynamics.} {\bf (A)} Representation of the
effective potential $V(y)$ for different values of the parameter $\beta$ . {\bf (B)}  The MFPTs of the first three shortest telomeres. {\bf (C)}  The MFPTs of the shortest and second shortest
telomeres for different thresholds $T$. The MFPTs are statistically separated for small $T$
(almost one standard deviation apart).  Parameters are $\beta=0.045$, $a=3\pm 1/2$.$L_0 = 1000$.}
\label{f:figure2}
\end{figure}

The solution of (\ref{diffKME}) is given by
\begin{align}
{\tau}_0(y)=&\,\frac{2}{\eps}\int\limits_{z_0}^y\exp\left\{\frac{2b}{\eps a}(z-z_0)
-\frac{2(a+b)b}{a^3\beta}
\log\frac{a^2\beta z+\eps b}{a^2\beta z_0+\eps b}\right\}\nonumber\\
&\,\times\int\limits_z^1\frac{
\exp\left\{-\ds\frac{2b}{\eps a}(u-z_0)\ds
+\ds\frac{2(a+b)b}{a^3\beta}
\log\ds\frac{a^2\beta u+\eps b}{a^2\beta z_0+\eps b}\right\}}{\ds\frac{a^2\beta bu+\eps b^2}{(\beta bu+\eps)b^2}}\,du\,dz.
\end{align}
The change of variables
$$v=\frac{2b(u-z_0)}{a\eps},\quad \zeta=\frac{2b(z-z_0)}{a\eps}$$
and the expansion of the denominator in \eqref{KMz0} give
\begin{align}
 \tau_0(y)=&\,\frac{\eps}{2}\int\limits_{0}^{2b(y-z_0)/a\eps}\frac{e^\zeta}{
(1+B\zeta)^A}\int\limits_\zeta^\infty
e^{-v}(1+Bv)^A\,dv\,d\zeta[1+O(\eps)],\label{tau0y}
\end{align}

where $$A=\frac{2(a+b)b}{a^3\beta},\quad B=\frac{a^3\beta}{a+b}.$$
The large $\zeta$ asymptotics
\begin{align*}
\frac{e^\zeta}{
(1+B\zeta)^A}\int\limits_\zeta^\infty
e^{-v}(1+Bv)^A\,dv\sim\sum\limits_{j=0}^\infty\frac{\Gamma(A+1)}{\Gamma(A-j+1)}
\left(\frac{B}{1+B\zeta}\right)^j
\end{align*}
and (\ref{tau0y}) give that for small $\eps$
\begin{align*}
{\tau}_0(y)\sim&\frac{\eps}{2}\int\limits_{0}^{2b(y-z_0)/a\eps}
\left[1+\frac{AB}{1+B\zeta}+\cdots\right]\,d\zeta\\
=&\frac{b}{a}(y-z_0)+\frac{A\eps}{2B}
\log\left(1+B\frac{2b(y-z_0)}{a\eps}\right)+\cdots.
\end{align*}
To conclude, the explicit solution of (\ref{diffKME}) has the
asymptotic representation for small $\eps$
\beq
{\tau}_0(y)=\frac{b}{a}(y-\zeta_0)+\frac{A\eps}{2B}
\log\left(1+B\frac{2b(y-z_0)}{a\eps}\right)+\cdots. \label{tau0asympt}
\eeq
It is clear from (\ref{tau0asympt}) that the derivatives of $ {\tau}_0(y)$ are
uniformly bounded for $y\geq z_0$ and $\eps>0$, so the Kramers-Moyal expansion (\ref{KMz0})
converges. It follows that the {unscaled MFPT $n_{1\to z_0}$ is given by
\begin{align*}
n_{1\to z_0}\sim\frac{\bar\tau_0(1)}{\eps},
\end{align*}
that is
\begin{align}
n_0(y)\sim \frac{b}{a\eps}(y-z_0)+\frac{A}{2B}
\log\left(1+B\frac{2b(y-z_0)}{a\eps}\right)+\cdots\,.
\end{align}
This analysis clarifies the first phase, which consists of noisy drifting to $z_0$. The second phase corresponds to escape from $z_0$ over the threshold $T$.}
\begin{figure}[http!]
\includegraphics[width=0.5\textwidth]{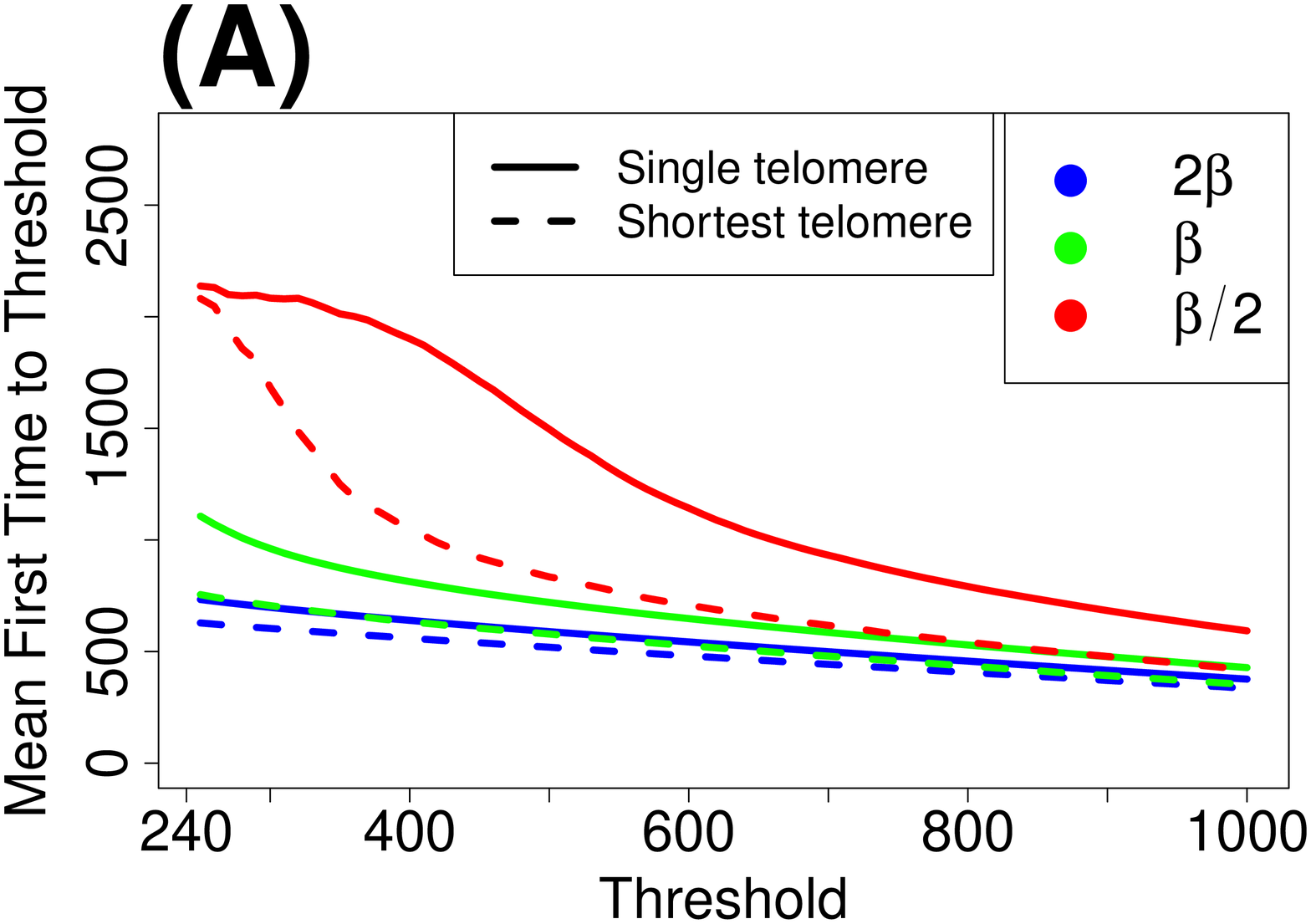} 
\includegraphics[width=0.5\textwidth]{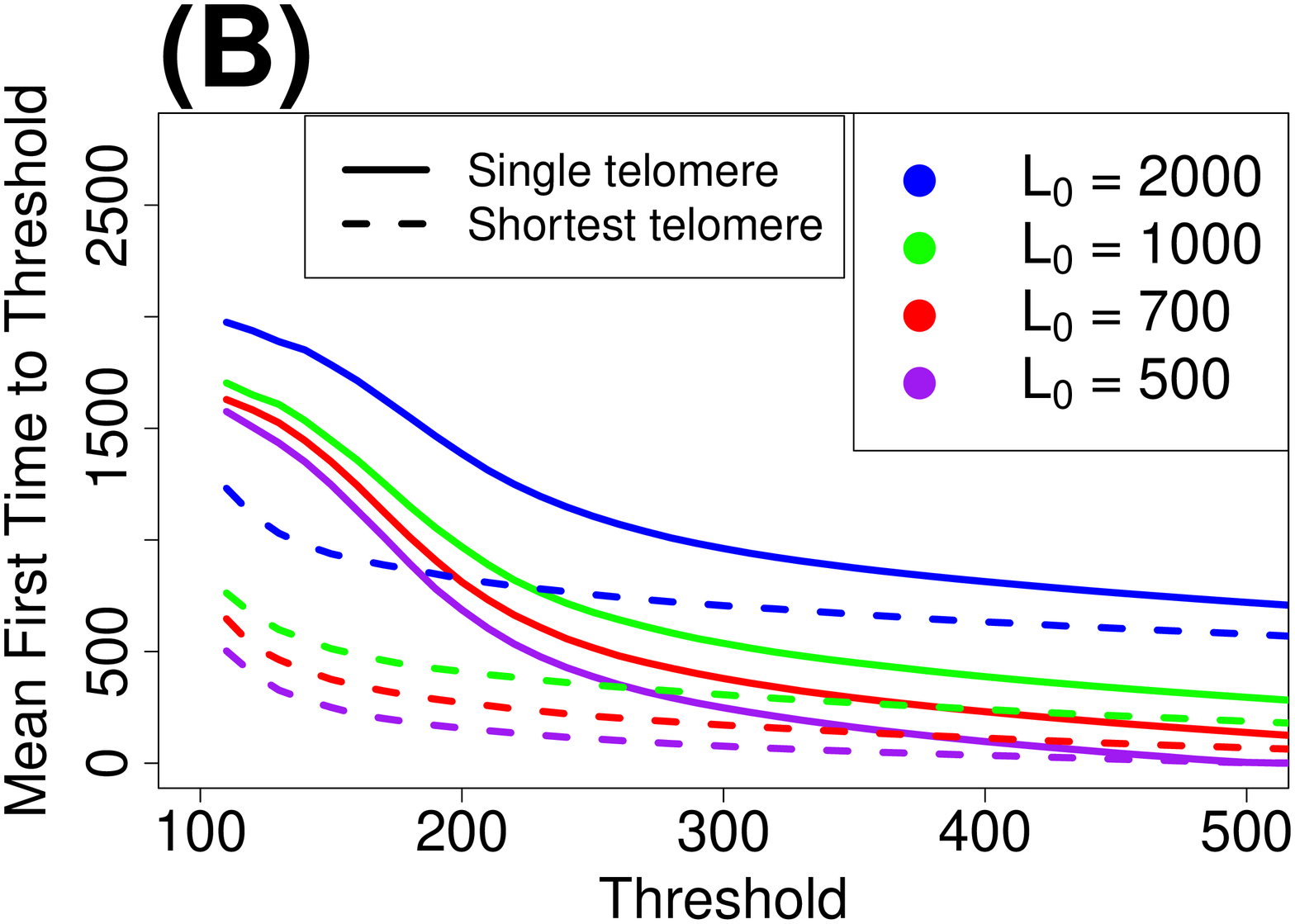}
\caption{{\bf Effect of changing the telomerase efficiency parameter $\beta$ on the MFPT to threshold.} \textit{\bf (A)} {The MFPT of a telomere and of the shortest among 32 telomeres} for $\beta$ (green),   $\beta/2$ (red) and $2\beta$ (blue). {\bf (B)} Effect of changing the initial length $L_0$.  Parameters are $\beta=0.045$, $a=3\pm 1/2$ and $L_0 = 2000$.}
\label{f:figure3}
\end{figure}

\begin{figure}[http!]
\includegraphics[width=1\textwidth]{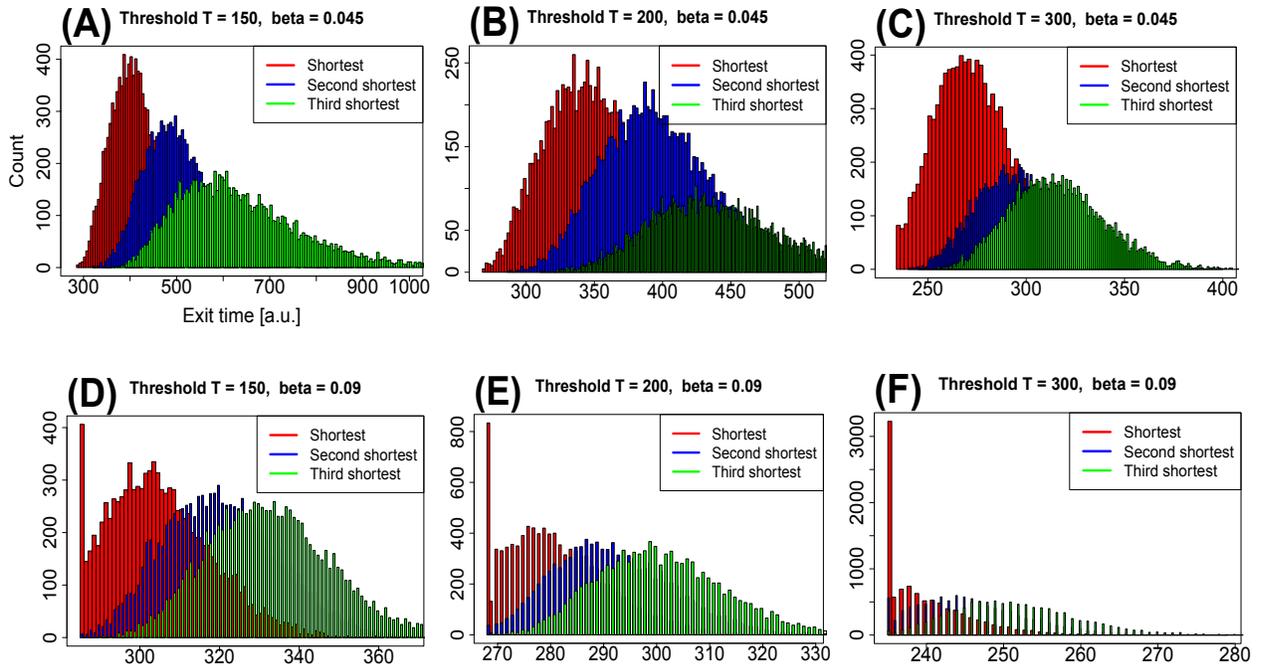}
\caption{{\bf Histogram of arrival time to a threshold T.} \textit{\bf (A,B,C)}
T=300,200,150. Parameters are $\beta=0.045$, $a=3\pm 1/2$ and $L_0 = 1000$. Number of runs = 10000.\textit{\bf (D,E,F)}:$\beta=0.09$ }
\label{f:figure4}
\end{figure}
\subsection{The role of the shortest telomere: time to senescence}
To further investigate the influence of telomere length distribution on the time to senescence, we resort to numerical simulations of the model \eqref{dynamics}. Interestingly, as the threshold decreases, the difference between the time to threshold increases, leading to a clear gap between the shortest and the second shortest lifetimes,  which is reduced between the second and the third.  This behavior can be interpreted by the two phases: as long as the threshold falls into the first phase,  the decay is deterministic and the difference between the first, second and third is insignificant. However, when the threshold is moved to the left of the critical point $z_0$ (e.g., $z=300$ for $\beta=0.045)$, reaching the threshold becomes noise-activation over a potential barrier, leading to a clear separation between the three}.

{To facilitate the calculation of the moments of the shortest and the second shortest lifetimes among $N$ i.i.d. trajectories of \eqref{dynamics}, we use the diffusion approximation \eqref{DA} to approximate the pdf of the first passage time to threshold. The solutions of
\begin{align}
p_t=&\,{\cal L}_np\quad\mbox{for}\ 0<y<1,\ p(y,0)=p_0(y),\ p(0,t)=p(1,t)=0\label{tdpdfn}\\
p_{\eps,t}=&\,{\cal L}_\eps p_\eps\quad\mbox{for}\ 0<y<1,\ p_\eps(y,0)=p_0(y),\ p_\eps(0,t)=p_\eps(1,t)=0,\label{tdpdfeps}
\end{align}
where the operators ${\cal L}_n$ and  ${\cal L}_\eps$ are defined in \eqref{BMEz0} and
\eqref{DA}, respectively, can be constructed by the method of separation of variable. The
dependence on time is exponential with exponents that are the eigenvalues of the boundary value
problems for the two operators. For small $\eps$, these are singular perturbation problems and
thus there is a big gap between the first and second eigenvalues in either case. The first
eigenvalue is the reciprocal of the MFPT \cite{Schuss}. Thus the eigenfunction expansions of the
pdfs are dominated by the first eigenfunction. Because the first eigenvalue is exponentially large in $1/\eps$, the pdf of survival time in the potential well can be approximated by a single exponential, which means that the survival time is Poissonian with mean $\tau_0(z_0)$. We use henceforward this approximation.}

We further investigate the gap between the first and the second MFPT by plotting in Fig.\ref{f:figure3}C  the mean and the variance.  We find that the standard deviations due to the shortest and to the second shortest overlap minimally, suggesting that the shortest telomere plays a key role in triggering senescence.  This result is due to the randomness of the model and depends on its parameters.

For $N$ Poissonian i.i.d. processes with escape time $\eE [\tau_1]=\bar\tau_1$, the expected shortest escape time is $\frac{\bar\tau_1}{N}$. The expected second shortest lifetime is $2\frac{\bar\tau_1}{N}$. Thus the gap between the first and the second is
\beq
\Delta=2\frac{\bar\tau_1}{N}-\frac{\bar\tau_1}{N}=\frac{\bar\tau_1}{N},
\eeq
which is the standard deviation of the first time $\eE [\tau_{first}]=\bar\tau_1/N$. Fig.\ref{f:figure4} indicates a deviation from the Poissonian case. Thus to study the effect of the shortest telomere, we use the ratio
\beq \label{ratio}
R=\frac{|\eE [\tau_{second}]-\eE [\tau_{first}]|}{\sqrt{\left( \eE [\tau_{first}^2]-\eE [\tau_{first}]^2 \right)}},
\eeq
where $\eE [\tau_{first}]$ (resp. $\eE [\tau_{second}]$) is the MFPT for the first (resp. second) telomere length to reach the threshold $T$. Interestingly, for a threshold $T=150$, we obtain that $R=1.92$ (the value of $\beta=0.045$ is given in table 1), while for the value $\beta=0.09$, we get $R=1.16$. This result suggests that decreasing the efficiency of the telomerase reduces the isolation of the shortest telomere relative to the second.
\begin{table}[!h]
	\begin{center}
		\begin{tabular}{lccccccc}
			Parameters   	             & Symbol       & & &Value & &  &  \\\hline
			
			Threshold 	         & T            & 150     & 150   & 200        & 200        & 300    & 300\\
			Beta & $\beta$ &0.045 & 0.09& 0.045 &0.09&0.045&0.09\\
			MFPT 1stshortest     & $\eE [\tau^T_{1,min}]$ &  410       & 305  & 348       & 281       & 274 & 239\\
			MFPT 2ndshortest     & $\eE [\tau^T_{2,min}]$ &  509   & 320  & 397      & 292     & 297 & 245\\
			MFPT 3rdshortest     & $\eE [\tau^T_{3,min}]$ & 640 & 333 & 442 & 301 & 316 & 251\\
			1st separation       &    $R_1$          & 1.92    & 1.16  & 1.42     & 1.07       & 1.17 & 1.09\\
			2nd separation & $R_2$      & 1.63     & 0.84   & 1.07  & 0.84     & 0.83   & 0.77
		\end{tabular}
	\end{center}
	\caption{Parameters estimated from the numerical simulation presented in figure \ref{f:figure4} .}
	\label{table:est_params}
\end{table}

\subsection{Effect of telomerase on cell lifetime}
To investigate the effect of telomerase on cell lifetime, we vary its efficiency by varying the parameter $\beta$ and the initial telomere length $L_0$. The results are shown in Figure \ref{f:figure3}: decreasing the parameter $\beta$  changes the position of the two phases and thus the time to threshold. We compare the mean and shortest among 32 telomeres arrival times to $T$ for three values of $\beta$ (green), $\beta$ (blue) and $\beta/2$ (red).  For the two latter ones, the mean and the shortest telomere length behave similarly  and deterministically. The deviation of the case $\beta/2$ (increase of the telomerase activity)  suggests that the time to senescence can be doubled relative to the normal case (Fig. \ref{f:figure3}A). Indeed, the quasi-steady state is shifted to the right (Fig. \ref{f:figure2}A). We conclude that the MFPT to threshold is not a linear function of $L_0$.  This effect of the telomere length diminishes as the threshold decreases (Fig \ref{f:figure3}B).

\section{Discussion and conclusion}
This paper analyzes telomere dynamics by an asymmetric random walk model of the length.
The asymmetry of the length is due to the decrease by a small amount $a$, compared to the large, but rare increase $b$ after each step of the walk (with state-dependent transition probability).  The main result of our model is that the time to senescence (measured by the length of the shortest telomere) is not proportional to the initial telomere length. Moreover,  the dynamics is divided into two phases: the first one is drift toward the quasi steady-state, where aging or the number of cell divisions is reflected in the telomere length. This could account for 30\% to 50\% of the number of divisions. In the second phase, the telomere length no longer reflects the number of divisions, but rather stays near an equilibrium length maintained close to the critical telomere size.  Due to random breaks and repairs, the telomere's length eventually reaches its critical size. The present stochastic model and computations are generic and can be applied to any cells.

Interestingly, the present analysis (Fig. \ref{f:figure1}B) reveals a clear difference between the mean and the shortest telomere length. Moreover, stochastic numerical simulations show a universal gap between the MFPT of the shortest and the second one (Fig. \ref{f:figure4}), suggesting that senescence can be triggered by the shortest telomere that reaches a given threshold. Any division, past this threshold would affect irreversibly the protein synthesis due to DNA. Because we do not know the exact threshold, we explore a continuum interval of values, but the results seem identical, showing a pronounced effect of a low threshold.

Although we previously studied the steady-state distribution of telomere length by the asymmetric random walk (model 1), we could not draw any conclusions about the dynamics, which requires a first passage time analysis of the shortest telomere. We found in \cite{Zhou,Daoduc} that there was a statistical gap between the distribution of the shortest and the second shortest telomere at steady state and that the gap depended on the parameters of the model.  Computing analytically the gap between the MFPT of the shortest and the second shortest to a threshold $T$ remains an open problem.

To study the consequences of telomere syndromes \cite{Armanios}, we varied the parameter $\beta$ that represents the telomerase efficiency: we found that increasing the value of $\beta$ by a factor 2, which is equivalent to decreasing the telomerase efficiency, has several consequences for senescence (Figs \ref{f:figure3} and \ref{f:figure4}): for example, with a threshold  of $T=150$, the time to senescence decreases from 410 to 305 (parameters in the figure legends). Another consequence of increasing $\beta$ is that the distribution of the shortest and the second shortest telomere get mixed, as shown above by the decrease of the ratio $R$ defined by \ref{ratio} from $R=1.92$ (for $\beta=0.045$) to  $R=1.16$ for $\beta=0.09$.

We used here a free parameter for the threshold $T$, which we varied, but in some known cases the minimum telomere length needed to ensure human telomere protective stability in white blood cells is 3.81 kb \cite{Blackburn2}. We conclude with general consequences of telomere shortening, which could be seen as a deregulation of time sensing: it is known that leukocyte telomere length can be used as a bio-marker of cardiovascular diseases, confirming that the distribution of telomere length and probably that of the shortest, plays a key role \cite{Blackburn2}. Stress hormones such as cortisol, is roughly inversely proportional to Leukocyte telomere length in a normal group of persons, but not in those suffering from Major Depressive Disorder \cite{Fair}. Similarly, adults suffering from major depression have shorter telomere length \cite{Blackburn2}. Finally, with aging,  the average telomere length decreases and this is correlated with an increase in mortality. Thus the measure of telomere shortness is also a statistical indicator of human mortality. To conclude, all these conditions can now be incorporated in modeling so that the distribution of telomere lengths and the shortest one can be predicted.


\end{document}